\documentclass[a4wide,10pt]{article}

\usepackage[round]{natbib}
\usepackage{graphicx}
\usepackage{amssymb}
\begin{document}


\title{Seismic Signal from Waves on Titan's Seas}

\author{Simon C. St\"ahler\thanks{mail@simonstaehler.com}\\Institute for Geophysics, ETH Z\"urich, Z\"urich, Switzerland \and
       Mark P. Panning, Steve Vance, Sharon Kedar \\ Jet Propulsion Laboratory, \\California Institute of Technology, Pasadena, CA, USA \and
       C\'eline Hadziioannou \\ Institute for Geophysics, \\Centrum f\"ur Erdsystemforschung und Nachhaltigkeit (CEN), \\University of Hamburg, Germany \and
       Ralph D. Lorenz \\ The Johns Hopkins University Applied Physics Laboratory, \\Laurel, MD, USA \and
       Knut Klingbeil \\ Leibniz Institute for Baltic Sea Research Warnem\"unde (IOW), \\Rostock, Germany
       }
\maketitle
\begin{abstract}
Seismology is the main tool for inferring the deep interior structures of Earth and potentially also of other planetary bodies in the solar system. Terrestrial seismology is influenced by the presence of the ocean-generated microseismic signal, which sets a lower limit on the earthquake detection capabilities but also provides a strong energy source to infer the interior structure on scales from local to continental. Titan is the only other place in the solar system with permanent surface liquids and future lander missions there might carry a seismic package. Therefore, the presence of microseisms would be of great benefit for interior studies, but also for detecting storm-generated waves on the lakes remotely. We estimated the strength of microseismic signals on Titan, based on wind speeds predicted from modeled global circulation models interior structure. We find that storms of more than 2 m/s wind speed, would create a signal that is globally observable with a high-quality broadband sensor and observable to a thousand kilometer distance with a space-ready seismometer, such as the InSight instruments currently operating on the surface of Mars.
\end{abstract}

\section{Introduction}
\label{sec:introduction}
\begin{figure}
 \centering
 \includegraphics[width=0.8\textwidth]{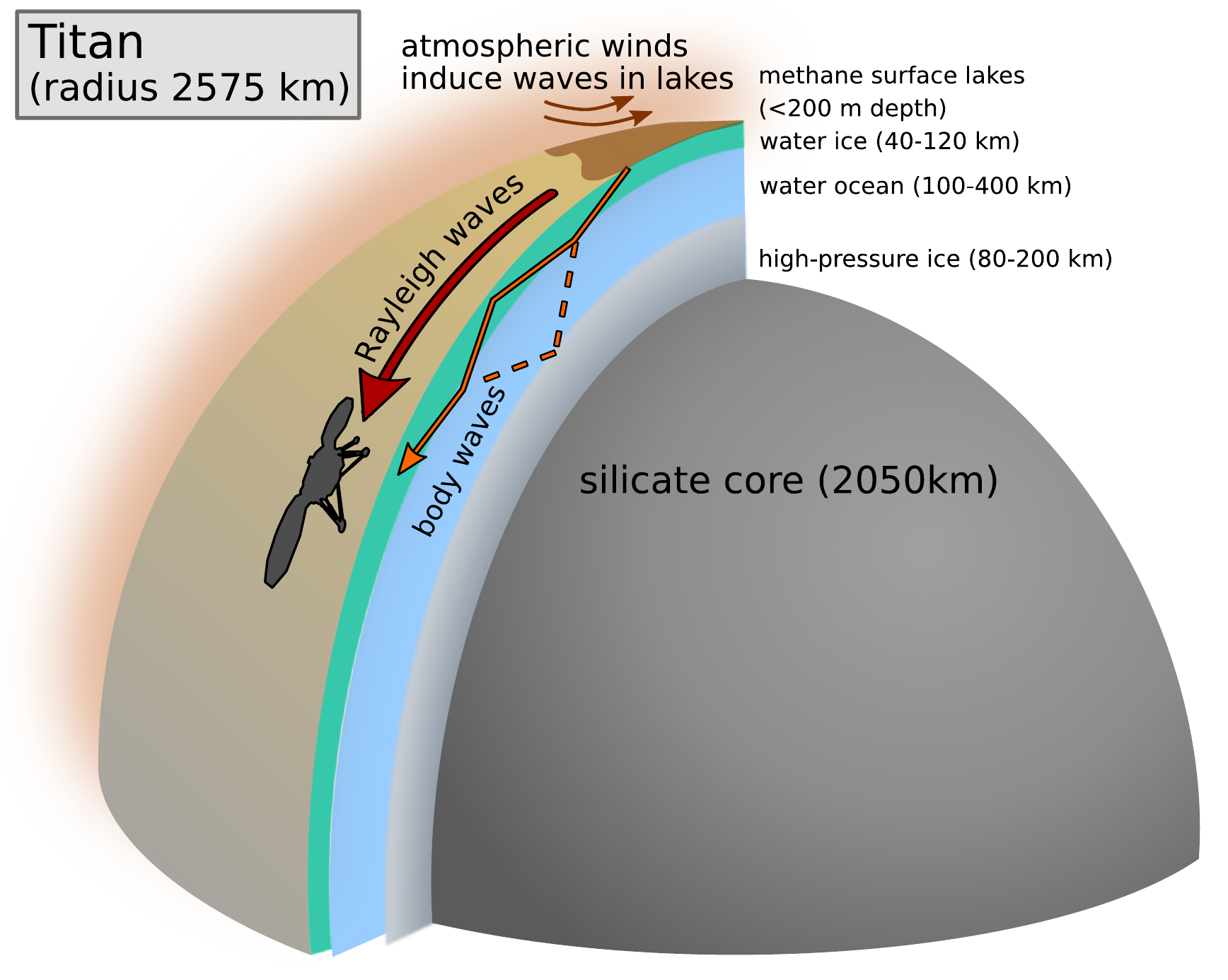}
 \caption{Illustration of the physical mechanism for generating microseisms. Winds in the atmosphere generate waves on the seas, which excite seismic waves within the ice shell or water ocean (body waves) or at the surface of the ice (Rayleigh waves). These waves could be measured over large distances by a potential seismometer-equipped lander---the graphical representation of which is not to scale.}
 \label{fig:overview}
\end{figure}
In many respects, Saturn's moon Titan is Earth's closest analog in the Solar System. It has a moderate-pressure atmosphere, a global seasonal climate system and liquid lakes on the surface \citep{Hayes2018}. On Earth, the combination of these three factors creates a globally observable seismic background signal, called ``microseismic noise'', due to its typical amplitude of a few micrometers per second \citep{Gutenberg1947}. This ``noise'' is the main limiting factor for seismic observations on Earth. At the same time, it is an invaluable signal source, because the noise is created by seismic waves that travel through and thereby sample the subsurface. Microseismic noise is employed by passive imaging methods to infer Earth's deep subsurface structure \citep{Shapiro2005, Boue2013} without having to rely on earthquake-- or artificially generated signals. Moreover, the microseismic signal is used to monitor temporal changes on all scales, from active faults \citep{Brenguier2008} to volcanoes \citep{Sens-Schonfelder2006}, down to the scale of engineering structures \citep{Snieder2006a}. Because the energy of these waves is related to ocean wave period and height, it carries a strong climate imprint, with the preferred locations moving from the Northern Atlantic and Pacific during Northern winter into the Southern Ocean during Northern summer \citep{Stutzmann2009}. Before the advent of weather satellites, it could even be used to detect offshore storms before they were observed on the coast \citep{Gutenberg1947}. Seismic wave generation through wave-wave interactions was theorized by \citep{Longuet-Higgins1950} and was confirmed using Wave Action Models by \citep{Hasselmann1963, Kedar2008a, Ardhuin2011}. 

The level of seismic background noise on Titan is unknown, but would be an important constraint for potential future seismometer-equipped landers, such as the Dragonfly concept \citep{Dragonfly2018}. For the Jupiter moon Europa, it has been estimated that the main seismic noise source is fracturing in its icy shell, created by tidal stresses \citep{Panning2018}. Compared to Europa, the tidal dissipation in Titan is smaller by approximately an order of magnitude \citep{Chen2014}, pushing the signal amplitude below the self-noise of space-ready seismometers. In this article, we estimate whether this means that Titan is completely quiet or whether the surface lakes (Maria) generate a persistent microseismic background (see fig. \ref{fig:overview}). Similar to Earth, such microseismic noise would limit the signal-to-noise ratio of titanquake observations, but more importantly, it could be used for passive imaging techniques and remote sensing of wave heights. As a reference, we use the SEIS instrument of InSight on Mars \citep{Lognonne2019}, which is currently the most sophisticated operational broadband seismic instrument. We start by reviewing the current models of wind-induced waves on Titan's lakes (section \ref{sec:wind_waves}) and transfer the terrestrial models of ocean microseism generation to Titan (section \ref{sec:excitation}) to finally arrive at a preliminary noise model for the surface of the moon under different wind scenarios (section \ref{sec:seismic_noise}), followed by a discussion of the modelling assumptions (section \ref{sec:discussion}). 

\section{Wind waves}
\label{sec:wind_waves}

\subsection{Winds}
Titan's lower atmosphere receives about 1000x less solar heating per unit area than the Earth, and the winds on Titan are comparatively gentle as a result. The Huygens probe recorded near-surface winds of less than 0.5 m/s, and the winds that cause sand to form the giant dunes that girdle Titan's equator are of the order of 1-1.5~m/s.  At higher latitudes, winds see strong seasonal forcing, with calm conditions in winter and fresher winds in the summer. At 90~m above Kraken Mare (Titan's largest lake) winds were estimated to be about 0.5 m/s for much of the year, rising to about 2 m/s during summer \citep{Lorenz2010}. At heights of 10~m, different Global Circulation Models (GCM) results \citep{Lorenz2012c, Hayes2013} suggested regular summer wind speeds at Kraken Mare and Ligeia Mare (Titan's second largest lake) of 1~m/s. It should be understood, however, that these are large-scale wind patterns determined by low-resolution GCMs, and may be thought of as comparable with the Earth's trade winds. Localized wind bursts may be considerably stronger.

It must also be noted that only isolated examples of possible waves on the surface of Titan's seas have been observed by Cassini \citep[e.g.]{Hofgartner2014, Barnes2014}, perhaps because of a slow freshening of winds in the northern summer before the end of the Cassini mission close to the solstice. One proposed explanation is that organic surface films inhibit wave creation \citep{Cordier2019}. However, the existence of shoreline erosion and deposition features, like the beaches on the North-western shoreline of lake Ontario Lacus \citep{Wall2010}, requires significant waves to occur at least occasionally.

Most likely, as on Earth, the strongest winds and the largest waves are associated with large rainstorms. These have been observed on Titan (e.g. near the south (summer) pole around Cassini's arrival in 2004, and near the equator during the equinox in 2009 \citep[e.g.][]{Turtle2011}). Cyclones are theoretically possible if the sea surface temperature rises enough \citep{Tokano2013}. Mesoscale models of methane rainstorms \citep{Barth2007} suggest that near-surface winds could reach 5~m/s within a few tens of km from a storm's center.

\subsection{Wave heights and periods}
\label{subsec:wave_heights}
The lakes on Titan's northern hemisphere are composed predominantly of methane and ethane \citep{Hayes2016}. While the physical parameters of liquid natural gas are different from those of water and the surface gravity $g=$1.354 m /s$^2$ is seven times smaller than on Earth, surface waves are in a similar regime: Waves of more than a few cm wavelength are pure gravity waves \citep{Srokosz1992} and capillary effects can be neglected. The dissipation of gravity waves depends on the depth of the liquid, i.e. on whether interaction with the ground has a considerable effect. In deep water, the dispersion relationship between wavelength $\lambda$ and period $T$ is controlled by gravity alone \citep[e.g.][]{Holthuijsen2007}: 
\begin{equation}
 \lambda=g T^2/\left(2\pi\right),
\end{equation}
which means that for a given period, wavelengths are about seven times shorter on Titan than on Earth. This implies that the minimum depth $d_{\mathrm{min}}$ for the deep-water approximation is less than 10 meter for all realistic wind conditions (see table \ref{tab:wavelengths}).

Under the condition of a fully developed sea, where the spectral input of energy from the winds is equal to the dissipation from breaking waves, the peak period of gravity waves $T$ depends linearly on wind speed $U_{10}$ at a height of 10 meters above the surface and the inverse of $g$:
\begin{equation}
T = 2\pi U_{10} / g \approx 4.64U_{10} ,
\end{equation}
resulting in wavelengths of
\begin{equation}
\lambda = 2\pi U_{10}^2/g \approx 4.64U_{10}^2
\end{equation}
These relationships are based on the assumption that the phase speed of the waves should not exceed the wind speed ($c_{\mathrm{peak}} \rightarrow U_{10}$, see \citet[note 6A]{Holthuijsen2007}), which is a reasonable assumption given current the understanding of wave generation in liquid methane by winds at 170 K.
From this, we expect the periods shown in table \ref{tab:wavelengths} for a given wind speed $U_{10}$. The table also shows the mininum depth $d_{\mathrm{min}}$ of the deep water approximation. Even the shallowest parts of Titan's larger lakes are deeper than 10 m~and the water depth does not have to be taken into account for wind speeds below 3-4 m/s.
\begin{table}
  \begin{center}
  \begin{tabular}{l|llll}
  $U_{10}$[m/s] & $T$ [s] & $\lambda$ [m] & $d_{\mathrm{min}}$ [m] & $F_{\mathrm{min}}$ [km]\\
  \hline
  1 &  4.64 & 4.64 & 0.74 & 17 \\
  2 &  9.28 & 18.6 & 2.95 & 67\\
  3 &  14.0 & 41.8 & 6.65 & 151\\
  4 &  18.6 & 74.2 & 11.8 & 268\\
  \end{tabular}
  \end{center}
  \caption{Dominant wave period $T$, wave length $\lambda$, minimum depth for the deep water approximation $kd>1$, i.e. $d_{\mathrm{min}}=\lambda/2\pi=U_{10}^2/g$ and minimum fetch length $F_{\mathrm{min}}$ for a fully developed sea for different wind speeds.}
  \label{tab:wavelengths}
\end{table}

The definition of a fully developed sea requires a certain fetch length, which describes the length scale on which the wind affects the sea, i.e. the leeward distance to the next coast. By comparison to terrestrial values, \citet{Ghafoor2000} found a minimum fetch (in km) of $F_{\mathrm{min}} \approx 16.8U_{10}^2$, above which the sea can be regarded as fully developed. This condition can be met for wind speeds of less than 4~m/s on Titan's maria Kraken (longest dimension 1170~km), Ligeia (500~km) and Punga (380~km) in the North and for slightly lower wind speeds also on Ontario Lacus (235 km) in the South.
\subsection{Wave spectra}
\begin{figure}
 \centering
 \includegraphics[width=0.4\textwidth]{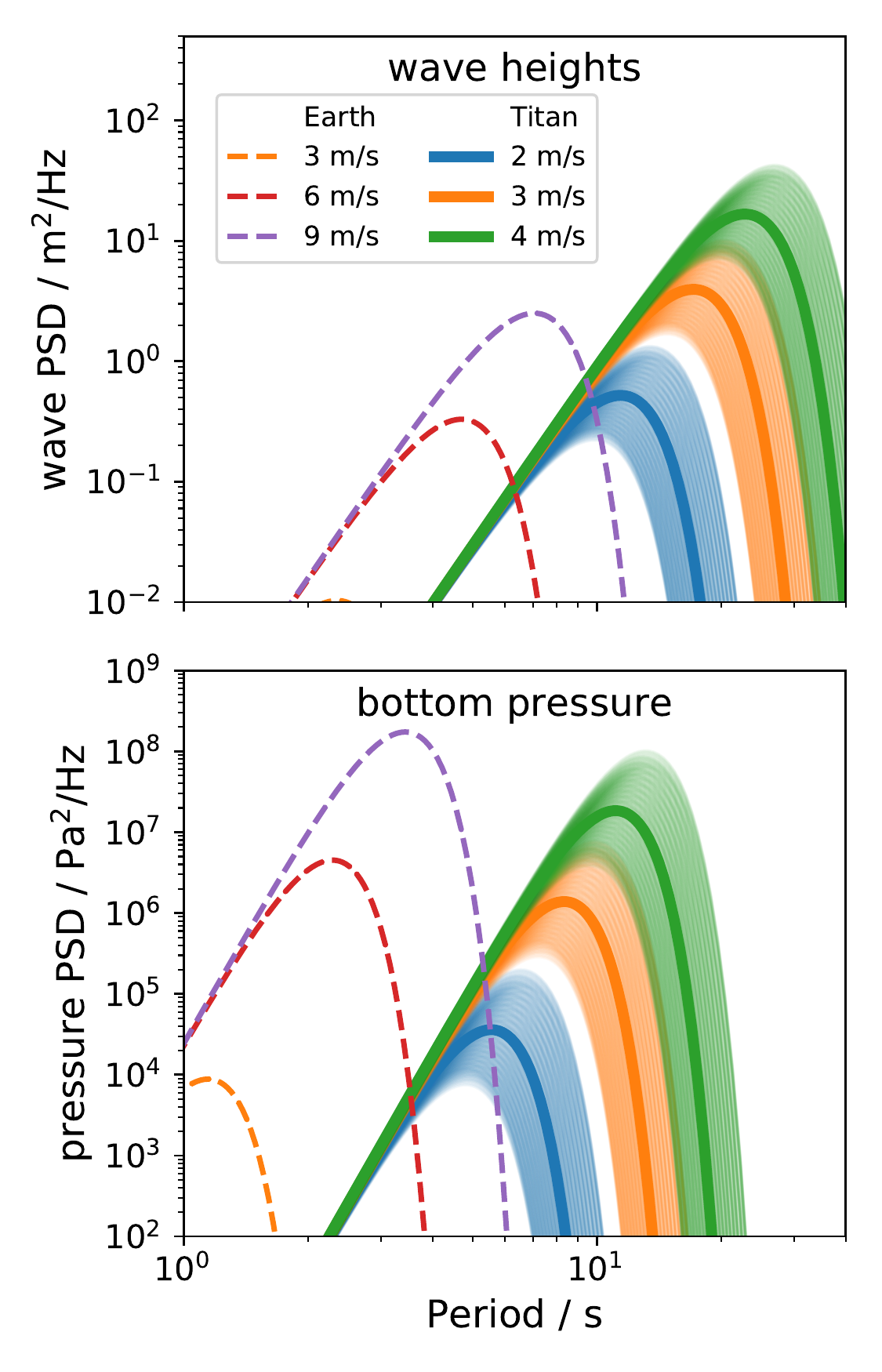}
 \caption{Wave height spectral density and resulting bottom pressure density compared between Earth and Titan. The band of values for Titan reflect a 20\% variation of the parameters $\nu, \alpha$ in the Pierson-Moskowitz model (eq. \ref{eq:wave height}).}
 \label{fig:spectra}
\end{figure}

From the previous arguments, we assume that we can work in the deep-water approximation and assume an unlimited fetch for a first estimate. To calculate the spectrum of seismic noise at a potential lander location, we need a model of the power spectrum of ocean waves. The wave spectrum $E(f)$ can be described by the Pierson-Moskowitz model \citep{Pierson1964}
\begin{equation}
E(f, f_m) = \frac{\alpha g^2}{(2\pi)^4 f^5}\exp \left( -\frac{5}{4} \left(\frac{f_m}{ f}\right)^4\right),
\label{eq:wave height}
\end{equation}
where $\alpha=8.1\cdot10^{-3}$ is the Philipp's constant. The dimensional peak frequency $f_m=\frac{\nu g}{U_{10}}$ relates the non-dimensional U-scaled peak frequency $\nu = 0.13$ to the surface gravity $g$ and $U_{10}$, the wind velocity at a height of 10 meters above the sea surface. We will continue to use this wind velocity in the further discussion.
The empirical coefficients $\alpha, \nu$ are probably slightly different on Titan, but following the discussion in \citet{Lorenz2012}, we assume that a model tested on Earth is better than a purely speculative model. Figure \ref{fig:spectra}A shows a comparison of the wave height spectra for the same wind speeds between Earth and Titan. A wind of 2 m/s will excite wave heights similar to a 6 m/s wind on Earth, but at three times the period. We allowed the parameters $\alpha, \nu$ to vary by 20\%, creating the bands in fig. \ref{fig:spectra}. The effect of the variation by 20\% is approximately that of a velocity change by the same magnitude.

\section{Excitation of seismic waves}
\label{sec:excitation}
Ocean waves generate seismic waves if two conditions are met:
\begin{enumerate}
 \item The temporal pressure variations at the ocean bottom must be high enough to generate an observable signal at seismic periods of a few seconds. Shorter periods will be dampened by attenuation, longer periods are difficult to observe with a small seismometer. The typical pressure variations necessary on Earth are of the order of a few kPa.
 \item The wave number of the pressure signal at the ocean bottom has to be similar to that of seismic waves of the same period \citep{Hasselmann1963}.
\end{enumerate}
\citet{Ardhuin2015} summarized the three situations under which these conditions are fulfilled for an ocean with depth $d$ and gravity waves with wave vector $\mathbf{k}$ that create seismic waves with wave number $k_{\mathrm{seis}}$:
\begin{description}
 \item[Primary Microseism on shallow bathymetry]: A gravity wave $\mathbf{k}_1$ in shallow water with $|\mathbf{k}|d<1$ causes a non-diminishing pressure signal at the seafloor. If the seafloor topography is periodic with wave vector $\mathbf{k}_b$, the resulting pressure can fulfill the condition
 \begin{equation}
   |\mathbf{k}_1+\mathbf{k}_b| = k_{\mathrm{seis}},
 \end{equation} 
 On Earth, this mechanism creates a signal at the dominant period of the terrestrial swell, roughly 14 seconds.
 \item[Long-period primary microseism]: Long-period ocean waves travelling toward a coastal slope create a pressure signal on the sea floor and excite seismic waves, if $|\mathbf{k}|d \approx 0.76$. Because this mechanism works for a large frequency range, it dominates the long-period seismic noise field ('hum') on Earth  at periods of 30--300~s. 
 \item[Secondary Microseism]: If two ocean waves with similar periods, $T_1$ and $T_2$, but almost opposite wave vectors $\mathbf{k}_1,\mathbf{k}_2$ interact, the resulting standing wave creates a pressure signal through the whole water column down to the ocean floor \citep{Longuet-Higgins1950}. If 
 \begin{equation}
   |\mathbf{k}_1+\mathbf{k}_2| = k_{\mathrm{seis}}  
 \end{equation}
 and $k_{\mathrm{seis}}^{-1} = (1/T_1+1/T_2) / c_R$ a Rayleigh wave is excited with group velocity $c_R = 0.87\beta$, $\beta$ being the shear wave velocity, and period $1/(1/T_1+1/T_2)$. This signal dominates at half the typical period of terrestrial swell, i.e. 7 seconds.
\end{description}

\subsection{Seismic parameters in Titan}
Titan's silicate interior is $\sim2110$~km in radius \citep{Vance2018}, covered by 460 to 550~km of water, resulting in a total average radius of 2575~km. This water is frozen at the uppermost 40 to 120~km (55 to 80 km, according to analysis of the Schumann resonance, see \citet{beghin2012analytic}). On top of the ice layer, lakes of liquid ethane and methane occupy much of Titan's northern hemisphere and some of its southern hemisphere. The seismic wave velocities $\beta$ in Titan's ice have been modeled in \citet{Vance2018} and are on the order of 2000~m/s. The anelastic attenuation is strongly temperature dependent and therefore very low at the surface (94~K, $Q\approx 1000$), but as the temperature increases rapidly within a few kilometers to 220 K, to stay at this value, $Q$ might be as low as 70 for much of the crust. Given the limited knowledge of $Q$ in very cold ices over a large range of pressures, there is a large uncertainty here \citep{Peters2012}.
For the sound speeds $\alpha$ of liquid ethane and methane, we used the values at 94~K from \citet{Younglove1987} (see table \ref{tab:seismic_values})
\begin{table}
\begin{tabular}{c|ccc}
     & density $\rho$ & sound speed $\alpha$ & shear wave velocity $\beta$ \\ \hline
ethane & 641 kg/m$^3$ & 1945 m/s & - \\ 
methane & 447 kg/m$^3$ & 1506 m/s & - \\
water ice & 932 kg/m$^3$ & 3878 m/s  & 1961 m/s \\
\end{tabular}
\caption{Seismic parameters used in this study. Ontario Lacus was modeled as pure ethane, the Northern maria were assumed to be pure methane.}
\label{tab:seismic_values}
\end{table}

The composition of some of the maria and larger lakes has been estimated from their radar absorption: Ontario Lacus is assumed to be mainly ethane \citep{Hayes2010}, Ligeia Mare is mainly methane \citep{Mitchell2015}, while for Kraken the composition is still unknown. A notable difference from Earth is that the sound speed in the surface liquids $\alpha_1$ is equal or only slightly lower than the shear wave speed $\beta_2$ in the icy crust.

\subsection{Secondary microseism}
\begin{figure*}
 \centering
 \includegraphics[width=0.8\textwidth]{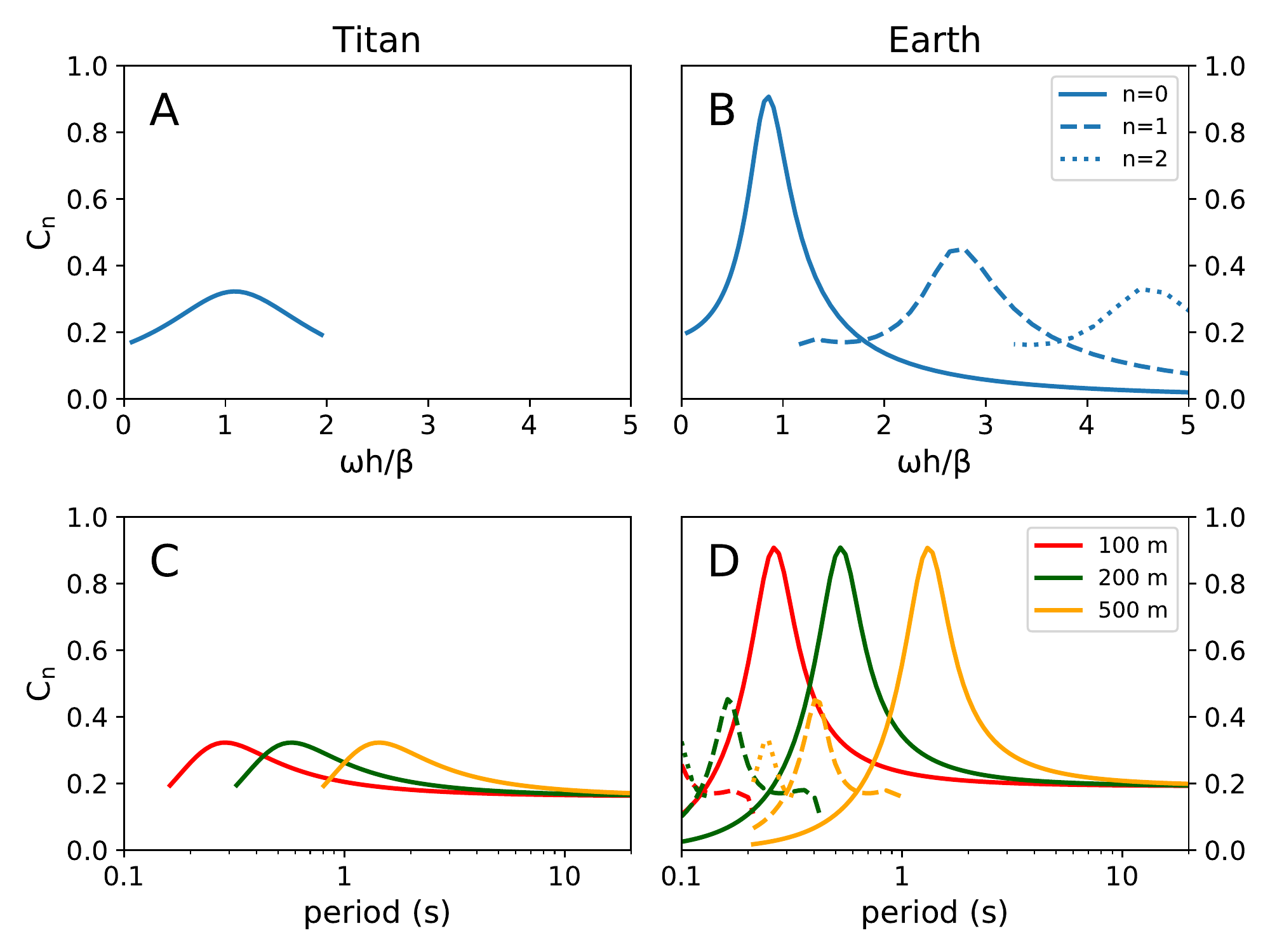}
 \caption{Amplification factor for Rayleigh waves. The left column shows values for Titan, the right one for Earth. The top row shows the amplification term as a function of the dimensionless frequency $\omega$ times depth $h$ divided by shear wave velocity in the lower medium $\beta_2$. Plot B is equivalent to fig. 2 in \citet{Longuet-Higgins1950}. The lower row shows the values for the specific depths in a methane lake above a water ice-crust (C), and a water ocean above a silicate crust (D).}
 \label{fig:amp_fac}
\end{figure*}
The secondary microseism mechanism requires the fewest assumptions on the bathymetry of the lakes, so we discuss it here in more detail:
\begin{enumerate}
\item 
Following \citet{Hasselmann1963, Ardhuin2011}, the bottom pressure spectrum for an ocean of density $\rho_1$ is 
\begin{equation}
F_{p3D}(f_2=2f) = \rho_1^2 g^2 f_2 E^2(f) I(f).
\label{eq:bottom_pressure}
\end{equation}
Note that the frequency of excited seismic waves $f$ is half the frequency of the ocean waves ($f_2$). $I(f)$ is a non-dimensional function that depends on the directional wave energy distribution $M(f, \theta)$, 
\begin{equation}
I(f) = \int_0^{\pi} M(f, \theta) M(f, \theta+\pi) d\theta.
\end{equation}
This directional term contains the requirement that was stated by \citep{Longuet-Higgins1950} and \citep{Hasselmann1963} regarding almost precisely opposite wave trains. Because we want to estimate an upper limit of the seismic noise density, we set $I(f) = 0.5$. This is the maximum value because $M(f,\theta)$ is normalized over $\theta$. 
Figure \ref{fig:spectra}B shows a comparison of the pressure spectra for the same wind speeds between Earth and Titan. A 3 m/s wind will excite two orders of magnitude higher pressure power, at seven times the period.

\item 
The ocean bottom pressure $F_{p3D}$ results in a vertical ground displacement power of 
\begin{equation}
S_{DF}(f_s) = \frac{2\pi f_s C(f_s, d)}{\rho_2^2 \beta_2^5}  F_{p3D}(f_2=f_s)
\label{eq:source_displacement}
\end{equation}
 at the source. The term $C(f_s, d)$ describes the displacement from Rayleigh waves at the lake-ice interface by an effective pressure source on the ground and takes resonance within the water column into account. 
For the seismic parameters chosen by Longuet-Higgins, and for more realistic earth models, this pressure results in a strongly frequency-dependent excitation of Rayleigh waves \citep{Kedar2008a, Gualtieri2013}. As shown in Fig.~\ref{fig:amp_fac}, the excitation function is relatively flat for seismic parameters of Titan's lakes---$\alpha_1 = 1420~ \textrm{m/s}, \rho_1=660~\textrm{kg/m}^3$ for the liquid layer and $\alpha_2 = 4000~\textrm{m/s}, \beta_2 = 2000~\textrm{m/s}, \rho_2=900~\textrm{kg/m}^3$ for the solid ice below. Unlike on Earth, the excitation of surface wave overtones is not possible on Titan.

In the shallow lakes of Titan, $C(f)\approx 0.32$ for the fundamental mode (n=0) at periods between 1 and 10~s.
The excitation spectrum of Rayleigh waves is therefore only controlled by the ocean bottom pressure spectrum created from the surface gravity waves and the depth profile can be neglected.

\item In a great circle arc distance $\Delta$, the spectral displacement power of the source described by (\ref{eq:source_displacement}) is
\begin{equation}
  F_{\delta}(f_s) = \frac{S_{DF}(f_s)}{R \sin \Delta} 
	\exp\left(-\frac{2\pi f_s \Delta R} {v_gQ}\right) A,
\label{eq:dist_displacement}
\end{equation}
where $A$ is the area of excitation, $R$ the radius of Titan and $c_R\approx0.87\beta_2$ the group velocity of Rayleigh waves in the solid medium. For a distributed pressure source $S_{DF}(f_s, \lambda', \phi')$ at latitude $\phi'$, longitude $\lambda'$, the vertical ground displacement at receiver location $(\lambda, \phi)$ becomes
\begin{eqnarray}
  F_{\delta}(f_s, \lambda, \phi) &= & 
  \int_{-\pi/2}^{\pi/2} \int_{0}^{2\pi} \frac{S_{DF}(f_s, \lambda', \phi')}{ \sin \Delta(\lambda', \phi')} \nonumber \\
  & & \exp\left(-\frac{2\pi f_s \Delta(\lambda', \phi') R} {v_g Q}\right) R \sin \phi' \\
  & &\textrm{d}\lambda' \textrm{d}\phi' \nonumber
\end{eqnarray}
$\Delta(\lambda', \phi')$ is the great circle arc distance from $(\lambda', \phi')$ to the receiver.
\item 
For a first estimate of the possible microseismic noise power, we assume that all lake and sea surfaces have fully developed wave spectra and fulfill the Longuet-Higgins criterion for secondary microseisms, e.g. by reflection from shorelines. We use the radar-derived map of \citet{Lorenz2014} for the Northern seas and maria and picked the shoreline of Ontario Lacus from \citet{Hayes2016}.
\end{enumerate}

On Earth, the microseismic noise has a significant  component from body waves \citep[e.g.]{gerstoft2008,landes2010}. Their excitation is fundamentally similar to Rayleigh waves, but as  was shown by \cite{Gualtieri2014}, the excitation factors $C$ in eq. \ref{eq:source_displacement} are different. Specifically, body wave excitations are even more frequency-dependent than Rayleigh waves. We estimated the excitation factors for depths between 50 and 500~m and found that their maximum is generally way above 1 Hz. As stated above, periods of surface gravity waves on Titan's lakes tend to be  far above one second due to the low gravity (see table \ref{tab:wavelengths}). In this period range, the excitation coefficients are constantly 0.3 for P-waves and 0.07 for S-waves. This means that P-waves are excited roughly with the same efficiency as Rayleigh waves.

\subsection{Primary microseism}
Primary microseism excitation is harder to assess, given the very limited  bathymetry for Titan's lakes and seas. On Earth, the excitation is facilitated by 1.) the extended shallow regions of the continental shelves, where the water depth is similar to the wave length of ocean gravity waves and 2.) the existence of long-period infragravity waves.
From the limited knowledge about the bathymetry of Titan's lakes \citep{Mastrogiuseppe2014}, the depth profiles seem to be steep. Even assumeing a linear slope, as per \citet{Lorenz2014}, only a very limited area of the surface of the maria is shallow enough that the bottom pressure is non-vanishing. The smaller lakes on the other side are so small and shallow that higher waves will not form in the first place.

The long-period primary microseism that creates Earth's hum requires the existence of infragravity waves. On Earth, these waves are formed by energy transfer from overlapping wind waves at shorter periods. This energy transfer is relatively slow and therefore requires permanent excitation of surface gravity waves. The fact wind waves were  not observed during any of the Cassini passes does not rule their formation, but makes it less likely that persisent enough to  excitate infragravity waves.

For the reasons stated in the previous section, we focus our quantitative analysis on the excitation of secondary microseism as described by \citet{Longuet-Higgins1950}. 

\section{Background noise}
\label{sec:seismic_noise}
\subsection{Huygens lander location}
\begin{figure}
 \centering
 \includegraphics[width=0.4\textwidth]{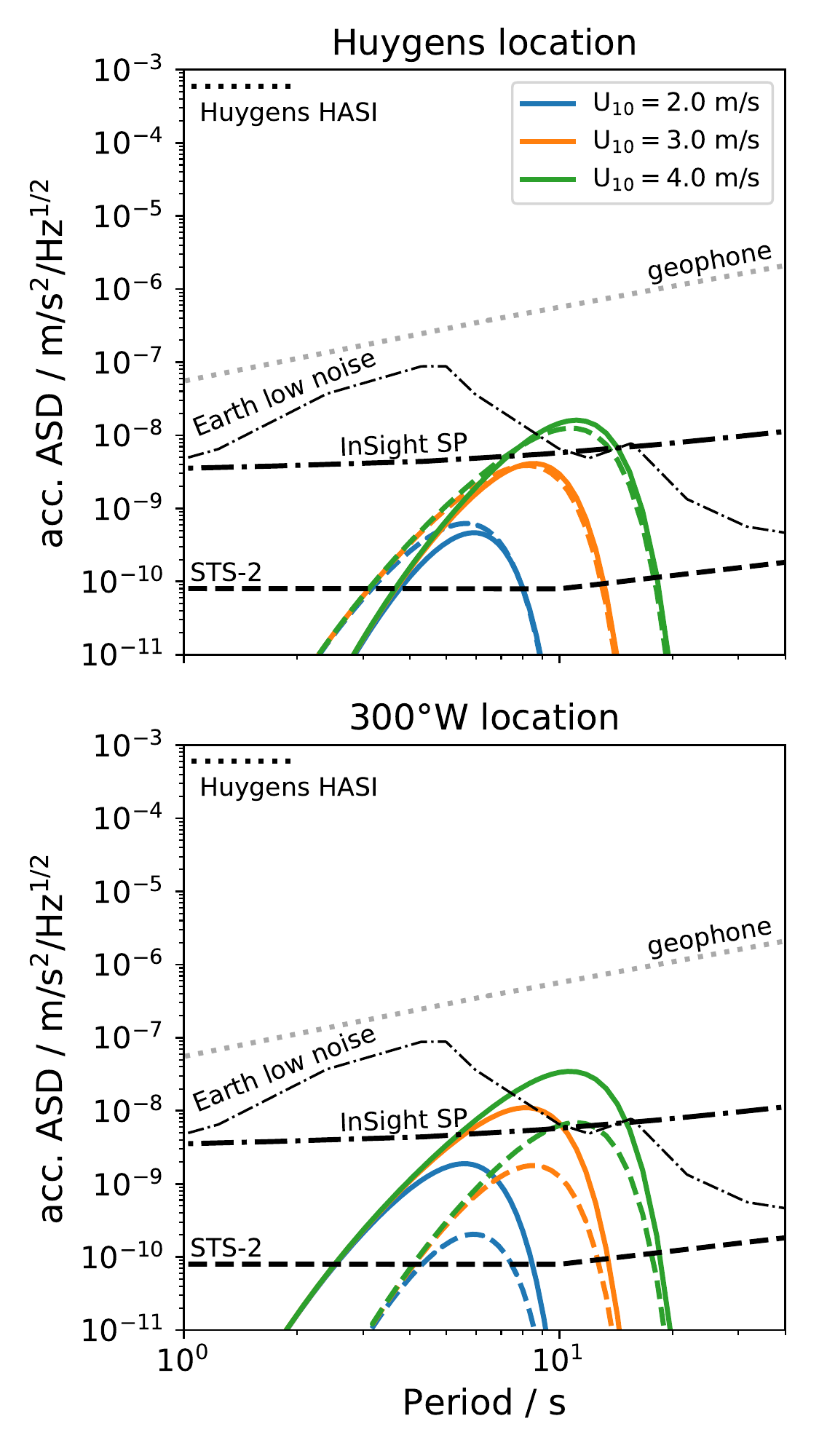}
 \caption{Microseismic vertical noise level for fully developed sea on Kraken Mare (solid, colored) or Ontario Lacus (dashed, colored) as recorded on the Huygens landing site (top) or a landing site in north-western Senkyo at 300$^\circ$W, 10$^\circ$N (bottom). The black curves are the self-noise level of an STS-2 seismometer (dashed), the InSight SP seismometer (dash-dotted), a passive geophone (dotted, grey) and the Huygens HASI accelerometer (dotted). The Earth low noise curve is the lowest noise usually observed at terrestrial stations \citep[New Low Noise Model, see][]{Peterson1993}}.
 \label{fig:noise_huygens}
\end{figure}

The Huygens entry probe landed on Titan in 2005, carrying an accelerometer (HASI) on board, which worked for 1200 seconds after landing \citep{Hathi2009}. The resolution of the sensor was $6\cdot10^{-4}$ m/s$^2$, and it picked up a number of spikes, probably from digital noise. \citet{Lorenz2017} discuss the signal level in the context of other planetary seismometers and accelerometers. 
Using the assumptions from the previous section, we estimated the microseismic noise spectrum at the Huygens landing site created by a fully developed sea of different wind speeds on Kraken Mare in comparison to Ontario Lacus (fig. \ref{fig:noise_huygens}). The results show that even the signal of a 4~m/s storm would be 100~dB below the resolution of the HASI sensor. Had Huygens carried the SP seismometer of the InSight mission \citep{Lognonne2019}, it would have been able to record this signal from either of the two seas. An STS-2 or the InSight VBB seismometer would have  been able to detect even a 2~m/s storm at this location.

\subsection{Landing locations for the 2030 decade}
Future missions to Titan could comprise many different architectures. NASA Flagship-class missions with an orbiter and one or more in-situ elements (landers, balloons etc...) could be relatively unrestricted in the place and time at which seismic measurements could be made, owing to the orbiter communication relay from the lander. However, single-element landed missions, such as those affordable in NASA's Discovery program, or in ESA's 'Medium-Class' category, would need to rely on Direct-to-Earth (DTE) communication. Unfortunately, a mission to the north polar seas is not possible after about 2024 \citep{Lorenz2015}, because the subsolar (and subearth) latitude is too far south and thus the Earth is never above the horizon as seen from Ligeia (78$^\circ$ N). Such missions become  possible again with continuous or near-continuous DTE in the next northern summer season, around the year 2040. 

A DTE mission to land near the equatorward margins of Kraken Mare, which reaches down below 60 $^\circ$N, could  be executed even during southern summer. However, with DTE only possible during parts of the day, such a mission would be perceived as technically risky. The terrain around Ontario Lacus (78~$^\circ$ S) is  well mapped in both the near-infrared and radar and the location would be favorable for DTE throughout the 2030s. Thus, although Ontario Lacus is much smaller in extent and so less likely to have strong microseismic activity, it could be a feasible landing site.

At the time of this writing, a relocatable lander mission concept, 'Dragonfly' \citep{Dragonfly2018} is being evaluated for Titan, with a competitive selection in 2019 in NASA's New Frontiers mission line. If selected, it would launch in 2025 and soft-land in 2034 using eight rotors and conduct measurements at a number of sites.  The principal objectives are astrobiological, but like the Viking landers at Mars, this in-situ chemical sampling would be supplemented by imaging, meteorology and geophysical measurements, including seismological observations. Information from seismic measurements is also astrobiologically important for inferring heat flux and interior chemistry that constrain Titan's habitability \citep{Vance2018,Vance2018a}. The initial landing site (in 2034) would be in interdune plains at low-moderate latitudes on Titan where safe landing sites are assured, but during a mission lasting several years such a mission could approach the seas.

As an example relevant to Dragonfly, we have evaluated the noise level at the Huygens landing site and at a purely hypothetical landing site in north-western Senkyo at 300$^\circ$W, 10$^\circ$N (fig. \ref{fig:noise_huygens}). At this position, a 3~m/s storm in Ontario Lacus would be well-detectable with the InSight SP-seismometer, while a storm on Kraken Mare would still have to have sustained winds faster than 4~m/s to create detectable seismicity. 

\subsection{Noise maps}
\begin{figure*}
 \centering
 \includegraphics[width=0.8\textwidth]{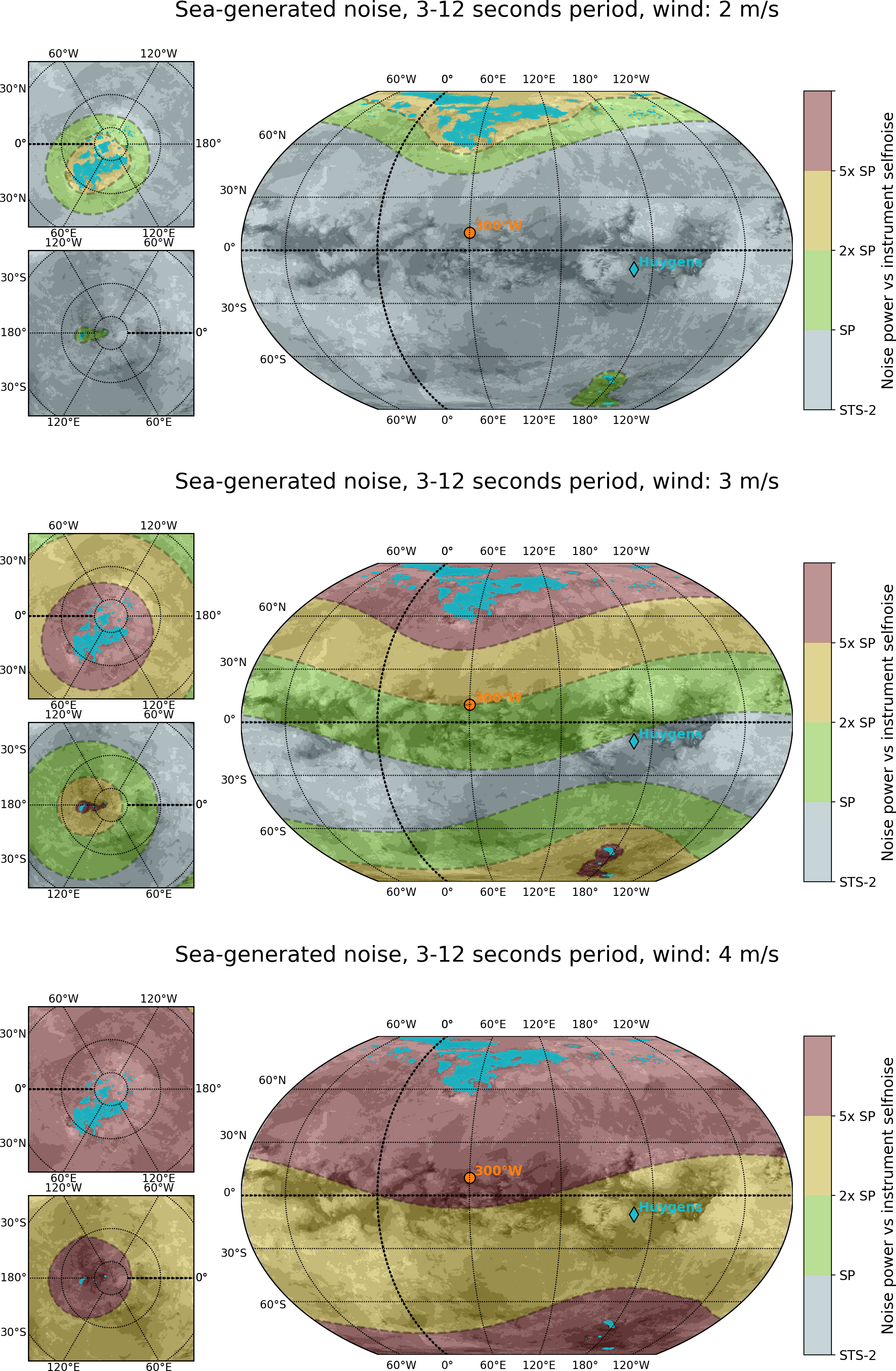}

 \caption{Map of sea-generated microseismic noise, assuming winds of 2 m/s (top), 3 m/s (center) or 4 m/s (bottom) had been blowing long enough to develop full seas on all liquid surfaces. The colorscale compares the noise to the self-noise of the InSight SP seismometer.}
 \label{fig:map}
\end{figure*}
To evaluate the detectability of microseismicity over the whole surface of Titan, we calculated the signal-to-noise ratio between the microseismic signal and the different instrument self-noises in a period range between 3 and 12 seconds (fig. \ref{fig:map}). The map shows that north of the equator, an SP-like instrument should be able to detect a 3~m/s storm everywhere, while the relatively small lakes of the Southern hemisphere restrict this range considerably. A 2~m/s storm would only be detectable much closer to the actual sea surfaces.

\section{Discussion}
\label{sec:discussion}
The arguments in this article are based on several assumptions. The strongest of them is that the global circulation models predict realistic wind distributions and, thus, that the lack of wave observations by Cassini RADAR passes was just bad timing. This question will probably not be settled until another spacecraft approaches Titan and measures wind speeds in situ. Given the known solar energy flux at Titan and the observed methane rain over large areas \citep{Turtle2011}, the atmosphere has to have stronger winds than the few observations of calm surfaces of the lakes suggest.

As discussed above, the short wavelengths of surface gravity waves on Titan mean that almost all waves in the lakes, if they occur, are deep waves with diminishing pressure at the bottom. This means that only the secondary microseism excitation mechanism will produce seismic waves, which require wave trains with opposite vectors. As discussed by \cite{Ardhuin2011}, opposite wave vectors are either a result of
\begin{enumerate}
 \item storms moving faster than the waves they create,
 \item two active storms in different parts of the sea,
 \item wave reflection from the shoreline.
\end{enumerate}
With the limited knowledge on storms over Titan, it is hard to determine the probability of mechanisms 1 and 2. \citet{Tokano2013} discussed that tropical cyclones could be possible over the polar seas. Such cyclones would most likely move fast enough for scenario 1. On the other hand, they would be too unlikely to assume that two could be active at the same time, as necessary for scenario 2. The shorelines of the Northern Maria are fractal, which may result in ineffective reflection of waves and therefore exclusion of mechanism 3. The shorelines of Ontario Lacus in the South however are relatively straight, so that waves could be reflected more effectively there, allowing mechanism 3.

Our model of wave spectra is simple, following Pierson-Moskowitz (PM) only. In terrestrial oceanography, this model was enhanced by \citep{Hasselmann1976} based on data from the JOint North Sea WAve Project (JONSWAP), which takes into account the finite fetch. 
For fetch lengths above a few km, this model results in a more pronounced peak than the original PM model, as observed in oceanographic data on Earth. 
Given that this model has several empirical terms, which have been found to differ from place to place on Earth, we adopted the simpler PM model, which is more conservative in maximum spectral energy. 

In our estimation, we assume that the wind wavefield is isotropic. \citet{duennebier2012} investigate the effect of directionality of wind waves, and estimate values of the overlap integral $I$, based on observed wind wave spectrum and ocean bottom acoustic observations.
They find values of up to $I$ = -20~dB for high wind speeds ($15~$m/s) and low frequency waves ($<1~$Hz). For higher frequencies and lower wind speeds ($<10~$m/s), the values of the overlap integral range between -10 and 0 dB, indicating a wavefield that approaches isotropic conditions ($I$ = -8~dB).
When the assumption of a fully isotropic wind wave distribution does not hold, the value of $I$ is found to vary by 2 to 10~dB for wind and wave conditions comparable to those on Titan. Note that these values are obtained in the case of an open ocean, where coastal reflections play a negligible role. For the seas and lakes on Titan, we expect shore reflections to contribute to creating a more isotropic wavefield, therefore putting our $I$ at values close to those chosen in this study. 

If detectable microseismic noise exists on Titan, it constitutes an important signal source for measurements of the atmosphere. In the scenario of a seismometer-equipped lander without a concurrent orbiter of either Titan or Saturn, recordings of seismic noise would be the only remote-sensing option. Even with the limitations discussed above, a tropical cyclone with wind speeds above 2 m/s over any of the lakes should be observable by a seismic instrument comparable to the InSight SP on Titan's surface.

The study compared the obtained signal levels with the modelled self-noise of the InSight instruments. Preliminary results of the InSight mission on Mars show that in planetary environments like the surface of Mars, and presumably Titan, the noise level of a seismic installation is mainly controlled by the instrument itself \citep{Clinton2019, Pike2019}. One reason for this simple noise sourcing is that the sandy surfaces found on planetary bodies allows for better ground coupling than bedrock or soil on Earth. An instrument installed on the deck of a lander or inside it, would be much more susceptible to atmospheric perturbations. As it is shown in figure \ref{fig:noise_huygens}, the frequency of the microseismic noise is generally between 0.05 and 0.2~Hz and decreases for stronger winds. This means that a geophone, a passive instrument with sensitivity limited to frequencies above 1 Hz, would generally not be able to observe lake microseism under any conditions. However, the InSight SP instrument, with a ruggedness and size that is comparable to a geophone, could be just good enough to detect these signals.

As discussed by \citet{Stahler2018}, Titan's Rayleigh waves have a period range limited by the thickness of the ice shell. Waves with periods $T>2\pi d/c_R$ are affected by the solid-liquid interface at the ice bottom and must be described as flexural waves. For 30~km, the smallest ice thickness compatible with available data, this is about 90 seconds, which is much higher than all realistic wave periods, therefore microseism will be Rayleigh or P-waves. This means that the microseismic signal cannot be used directly to constrain the ice thickness, e.g. by analyzing its polarization. Body wave reverberations would have resonance periods of $T=2d/v$, which is about 14~seconds for P-waves (at a ice thickness of 28~km). Given a large storm that would excite a broad period range of seismic waves, this peak could be detected. Since the source spectrum of the seismic signal would be \textit{a priori} unknown, this analysis would not be unambiguous.
A scientific question that can almost only be answered by seismology, is the existence of high-pressure ice phases below the water ocean of Titan. These layers would create clear spectral peaks from S-wave multiples in the coda of Pn-waves of titanquakes for teleseismic distances around 40 degree. The microseismic signal itself would probably be too weak to answer this question.
The subsurface (water) ocean might have seiche-like eigenmodes on its own that are driven by convection. The signal shape of these modes was estimated by \citet{Panning2018} for the Jupiter moon Europa and found to be just above the self-noise of an STS-2 at periods of hundreds of seconds. With its thicker ice shell, Titan has less convection than Europa and therefore we assume that this signal will not be detectable with a space-ready seismometer. The same is true for potential tides and seiches of the surface lakes, which will be at periods of minutes to hours and therefore not observable by an unburied seismometer without dedicated shielding against temperature changes. 

A final question is whether ocean-generated microseismic noise would interfere with classical event-based seismology. Even in the scenario of a 4 m/s storm on either Kraken Mare or Ontario Lacus, the generated noise will be below the New Low Noise Model on Earth \citep{Peterson1993}, which is the empirical lower bound of noise observed at the stations of highest quality on Earth. Since it is below the self-noise of any realistic seismic instrument for most of the days, the effect of microseismic noise on the detection of titanquakes will be small.

\section{Conclusion}
\label{sec:conclusion}
We have shown  that detectable microseismic noise will not be a regular feature on Titan, but rather will be limited to a few days per year, due to the small size of Titan's seas and the relatively low wind speeds. This means that microseismic noise will not interfere with potential future seismic observations on Titan. In fact, only microseismic noise created by major storms over the seas of Titan would be detectable with a space-ready seismometer on much of the Northern hemisphere. Thus, a long-term installation of a seismometer on the surface of Titan would allow  monitoring of the occurrence of strong storms in the polar regions without the need for additional observations from orbit or a dedicated polar lander.

\section{Acknowledgements}
\label{sec:Acknowledgements}
CH acknowledges support from the Emmy Noether program (grant no. HA7019/1-1) of the Deutsche Forschungsgemeinschaft (DFG). RL acknowledges the support of NASA Outer Planets Research Grant NNX13AK97G "Physical Processes in Titan's Seas". Work by JPL co-authors was partially supported by strategic research and technology funds from the Jet Propulsion Laboratory, Caltech, and by the Icy Worlds and Titan nodes of NASA’s Astrobiology Institute (13-13NAI7\_2-0024 and 17-NAI8\_2-017). The work of KK was supported by the Collaborative Research Centre TRR 181 “Energy Transfer in Atmosphere and Ocean” (project number 274762653), funded by the DFG. Computations were done at the Swiss National Supercomputing Center (CSCS) under project ID sm682.

\appendix

\section{Attenuation}
Following \citet{Cammarano2006, Vance2018}, the attenuation is: 
\begin{equation}
    Q = B\omega^\gamma \exp\left( \frac{\gamma g_a T_m}{T} \right) 
\end{equation}
with B=0.56 a normalisation constant, $\gamma=0.2$, $g_a=22$ for iceI and $T_m=245$~K, the melting temperature. The surface temperature is 94~K, increasing to 240~K in the uppermost 7 km of the ice and then slowly to 245~K at the bottom of the ice. We evaluated the shear wave attenuation at $f = 1$~Hz, to arrive at value of $Q=70$ for most of the ice.

\section{Amplification factors on Titan and Earth}
\subsection{Rayleigh waves}
The amplification term $c_m$ in eq. \ref{eq:source_displacement} is calculated by equation 184 in \citet{Longuet-Higgins1950}
\begin{equation}
 c_m = (-1)^m\left(\frac{\beta_2}{\omega}\right) ^{5/2}  
    \frac{\sqrt{k_{0,m}}}{\partial G / \partial k|_{k_{0,m}}},
\end{equation}
which has to be evaluated at $k_{0,m}$, the roots of $G(k)$, (eq. 179 ibd):
\begin{eqnarray}
G(k)& = & \frac{\beta_2 ^ 4}{\omega ^ 4} 
\left( \frac{\left(2 k^2 - \omega^2/\beta_2^2\right)^2}{\sqrt{k^2 - \omega^2 / \alpha_2^2}}
                         - 4 k^2 \sqrt{k^2 - \frac{\omega^2}{\beta_2^2}} \right)
             \cosh\left(\sqrt{k^2 - \frac{\omega^2}{\alpha_1^2}} h\right) \nonumber \\
           & & + \frac{\rho_1/\rho_2}{\sqrt{k^2 - \omega^2/\alpha_1^2}} \
             \sinh\left(\sqrt{k^2 - \frac{\omega^2}{\alpha_1^2}}  h\right).
\end{eqnarray}
Function $G(k)$ becomes more oscillatory for higher ratios of shear wave speed $\beta_2$ in the lower medium to sound speed in the ocean $\alpha_1$. For the seismic parameters of Titan, it generally has only one root, corresponding to fundamental mode Rayleigh waves.

\subsection{Body waves}
\begin{figure*}
 \centering
 \includegraphics[width=0.95\textwidth]{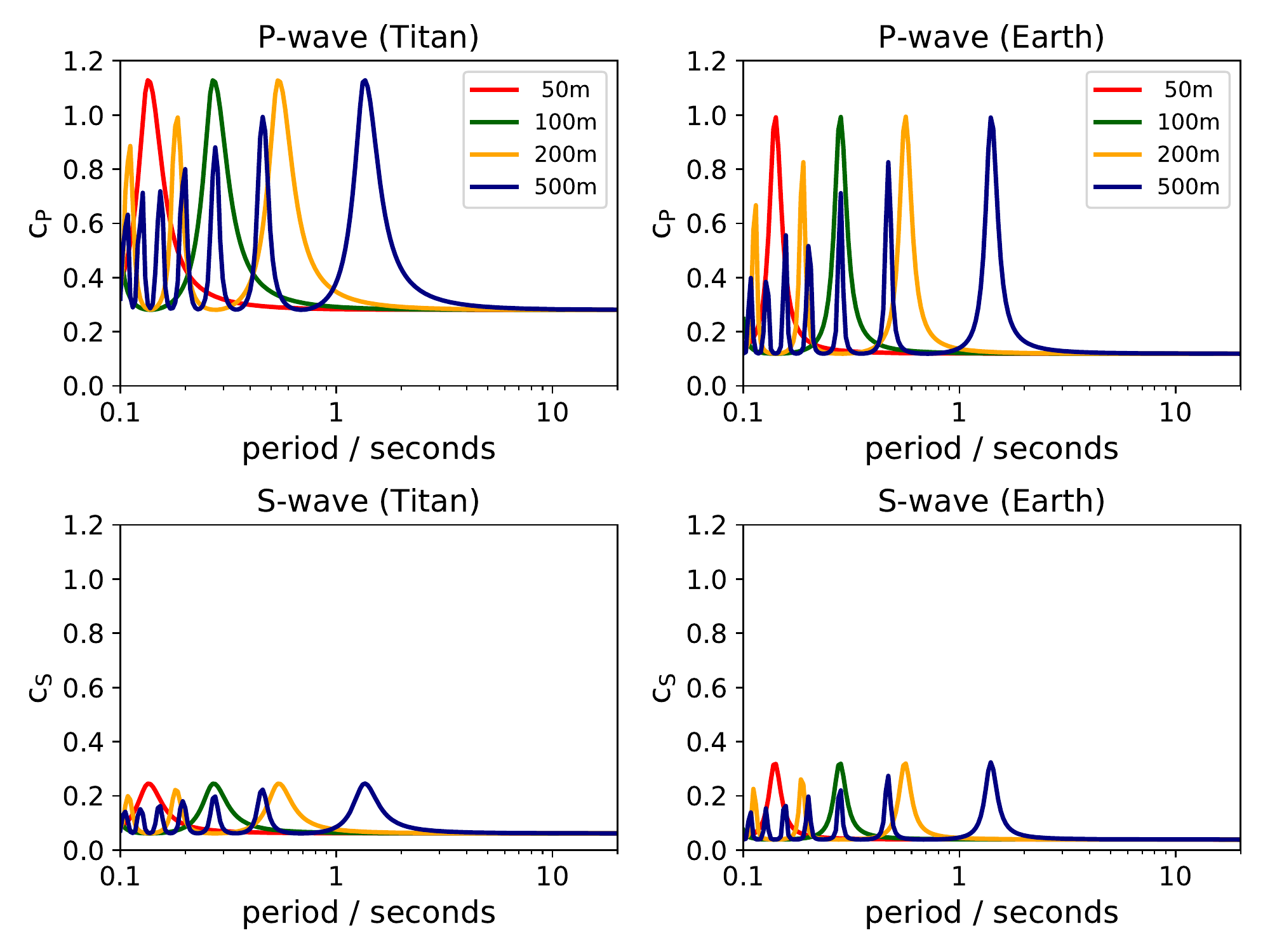}
 \caption{Amplification factor for body wave excitation. As on Earth, this mechanism is much more effective for P than for S-waves.}
 \label{fig:P_amplitude}
\end{figure*}

Following \citet{Gualtieri2014}, the amplification coefficients for body waves $c_P(h,\omega), c_S(h,\omega)$ can be calculated by integrating the contribution of all P-waves reflected in the water layer before they are being transmitted into the crust over all angles $\theta$ smaller than the critical angle for full reflection $\theta_{\mathrm{cr},X}$
\begin{equation}
 c_X(h,\omega) = \left(
    \int^{\theta_{\mathrm{cr},X}}_0 
    \left|
          \frac{T_X(\theta)}
               {1 + R_X(\theta)
                e^{i \Phi(h, \omega, \theta)}}
    \right|^2 \mathrm{d}\theta
    \right)^{1/2}.
\end{equation}
This value is complex and takes into account the phase shift $\Phi(h, \omega, \theta) = 2 \omega h / \alpha_1 \cos\theta$ obtained by the wave in the water column. $T_X$ and $R_X$ are transmission and reflection coefficients for the respective wave type at the sea floor.
As is can be seen in fig. \ref{fig:P_amplitude}, these amplification coefficients are highly oscillatory for periods shorter than one second, but fundamentally flat for longer periods.


\section{References}
\begin{thebibliography}{56}
\providecommand{\natexlab}[1]{#1}
\providecommand{\url}[1]{\texttt{#1}}
\expandafter\ifx\csname urlstyle\endcsname\relax
  \providecommand{\doi}[1]{doi: #1}\else
  \providecommand{\doi}{doi: \begingroup \urlstyle{rm}\Url}\fi

\bibitem[Ardhuin et~al.(2011)Ardhuin, Stutzmann, Schimmel, and
  Mangeney]{Ardhuin2011}
Fabrice Ardhuin, El{\'{e}}onore Stutzmann, Martin Schimmel, and Anne Mangeney.
\newblock {Ocean wave sources of seismic noise}.
\newblock \emph{J. Geophys. Res. Ocean.}, 116\penalty0 (9):\penalty0 1--21,
  2011.
\newblock \doi{10.1029/2011JC006952}.

\bibitem[Ardhuin et~al.(2015)Ardhuin, Gualtieri, and Stutzmann]{Ardhuin2015}
Fabrice Ardhuin, Lucia Gualtieri, and El{\'{e}}onore Stutzmann.
\newblock {How ocean waves rock the Earth: Two mechanisms explain microseisms
  with periods 3 to 300 s}.
\newblock \emph{Geophys. Res. Lett.}, 42:\penalty0 765--772, 2015.
\newblock \doi{10.1002/2014GL062782.1.}

\bibitem[Barnes et~al.(2014)Barnes, Sotin, Soderblom, Brown, Hayes, Donelan,
  Rodriguez, Mou{\'{e}}lic, Baines, and McCord]{Barnes2014}
Jason~W Barnes, Christophe Sotin, Jason~M. Soderblom, Robert~H Brown,
  Alexander~G. Hayes, Mark Donelan, Sebastien Rodriguez, St{\'{e}}phane~Le
  Mou{\'{e}}lic, Kevin~H Baines, and Thomas~B McCord.
\newblock {Cassini/VIMS observes rough surfaces on Titan's Punga Mare in
  specular reflection}.
\newblock \emph{Planet. Sci.}, 3\penalty0 (1):\penalty0 3, 12 2014.
\newblock \doi{10.1186/s13535-014-0003-4}.

\bibitem[Barth and Rafkin(2007)]{Barth2007}
Erika~L. Barth and Scot C.~R. Rafkin.
\newblock {TRAMS: A new dynamic cloud model for Titan's methane clouds}.
\newblock \emph{Geophys. Res. Lett.}, 34\penalty0 (3):\penalty0 L03203, feb
  2007.
\newblock \doi{10.1029/2006GL028652}.

\bibitem[B{\'{e}}ghin et~al.(2012)B{\'{e}}ghin, Randriamboarison, Hamelin,
  Karkoschka, Sotin, Whitten, Berthelier, Grard, and
  Sim{\~{o}}es]{beghin2012analytic}
Christian B{\'{e}}ghin, Or{\'{e}}lien Randriamboarison, Michel Hamelin, Erich
  Karkoschka, Christophe Sotin, Robert~C Whitten, Jean-Jacques Berthelier,
  R{\'{e}}jean Grard, and Fernando Sim{\~{o}}es.
\newblock {Analytic theory of Titan’s Schumann resonance: Constraints on
  ionospheric conductivity and buried water ocean}.
\newblock \emph{Icarus}, 218\penalty0 (2):\penalty0 1028--1042, 4 2012.
\newblock \doi{10.1016/j.icarus.2012.02.005}.

\bibitem[Bou{\'{e}} et~al.(2013)Bou{\'{e}}, Poli, and Campillo]{Boue2013}
P.~Bou{\'{e}}, Pierro Poli, and Michel Campillo.
\newblock {Teleseismic correlations of ambient seismic noise for deep global
  imaging of the Earth}.
\newblock \emph{Geophys. J. Int.}, 194\penalty0 (2):\penalty0 844--848, may
  2013.
\newblock \doi{10.1093/gji/ggt160}.

\bibitem[Brenguier et~al.(2008)Brenguier, Campillo, Hadziioannou, Shapiro,
  Nadeau, and Larose]{Brenguier2008}
Florent Brenguier, Michel Campillo, C{\'{e}}line Hadziioannou, Nikolai~M. Shapiro,
  R~M Nadeau, and Eric Larose.
\newblock {Postseismic relaxation along the San Andreas fault at Parkfield from
  continuous seismological observations.}
\newblock \emph{Science}, 321\penalty0 (5895):\penalty0 1478--1481, 2008.

\bibitem[Cammarano et~al.(2006)Cammarano, Lekic, Manga, Panning, and
  Romanowicz]{Cammarano2006}
Fabio Cammarano, V.~Lekic, M.~Manga, Mark~P. Panning, and Barbara Romanowicz.
\newblock {Long-period seismology on Europa: 1. Physically consistent interior
  models}.
\newblock \emph{J. Geophys. Res.}, 111\penalty0 (E12):\penalty0 E12009, dec
  2006.
\newblock \doi{10.1029/2006JE002710}.

\bibitem[Chen et~al.(2014)Chen, Nimmo, and Glatzmaier]{Chen2014}
E~M~A Chen, F~Nimmo, and G~A Glatzmaier.
\newblock {Tidal heating in icy satellite oceans}.
\newblock \emph{Icarus}, 229:\penalty0 11--30, 2014.
\newblock \doi{10.1016/j.icarus.2013.10.024}.

\bibitem[Clinton et~al.(2019)Clinton, Ceylan, St{\"{a}}hler, Giardini,
  B{\"{o}}se, van Driel, Horleston, Kawamura, Kedar, Khan, Euchner, Panning,
  Lognonn{\'{e}}, and Banerdt]{Clinton2019}
J.~F. Clinton, S.~Ceylan, S.~C. St{\"{a}}hler, D.~Giardini, M.~B{\"{o}}se,
  M.~van Driel, A.~Horleston, T.~Kawamura, S.~Kedar, A.~Khan, F.~Euchner,
  Mark~P. Panning, Philippe Lognonn{\'{e}}, and B.~Banerdt.
\newblock {Marsquake Service for InSight: Preliminary Observations and
  Operations in Practice}.
\newblock \emph{50th Lunar and Planetary Science Conference}, 50:\penalty0
  2915, 2019.
\newblock URL \url{http://adsabs.harvard.edu/abs/2019LPI....50.2915C}.

\bibitem[Cordier and Carrasco(2019)]{Cordier2019}
Daniel Cordier and Nathalie Carrasco.
\newblock {The floatability of aerosols and wave damping on Titan’s seas}.
\newblock \emph{Nat. Geosci.}, 2019.
\newblock \doi{10.1038/s41561-019-0344-4}.

\bibitem[Duennebier et~al.(2012)Duennebier, Lukas, Nosal, Aucan, and
  Weller]{duennebier2012}
Fred~K Duennebier, Roger Lukas, Eva-Marie Nosal, J{\'e}rome Aucan, and Robert~A
  Weller.
\newblock Wind, waves, and acoustic background levels at station aloha.
\newblock \emph{J. Geophys. Res. Oceans}, 117\penalty0 (C3),
  2012.

\bibitem[Gerstoft et~al.(2008)Gerstoft, Shearer, Harmon, and
  Zhang]{gerstoft2008}
Peter Gerstoft, Peter~M. Shearer, Nick Harmon, and Jian Zhang.
\newblock {Global P, PP, and PKP wave microseisms observed from distant
  storms}.
\newblock \emph{Geophys. Res. Lett.}, 35\penalty0 (23), 2008.

\bibitem[Ghafoor et~al.(2000)Ghafoor, Zarnecki, Challenor, and
  Srokosz]{Ghafoor2000}
Nadeem A.-L. Ghafoor, John~C Zarnecki, Peter Challenor, and Meric~A Srokosz.
\newblock {Wind-driven surface waves on Titan}.
\newblock \emph{J. Geophys. Res. Planets}, 105\penalty0 (E5):\penalty0
  12077--12091, may 2000.
\newblock \doi{10.1029/1999JE001066}.

\bibitem[Gualtieri et~al.(2013)Gualtieri, Stutzmann, Capdeville, Ardhuin,
  Schimmel, Mangeney, and Morelli]{Gualtieri2013}
Lucia Gualtieri, El{\'{e}}onore Stutzmann, Yann Capdeville, Fabrice Ardhuin,
  Martin Schimmel, A.~Mangeney, and Andrea Morelli.
\newblock {Modelling secondary microseismic noise by normal mode summation}.
\newblock \emph{Geophys. J. Int.}, 193:\penalty0 1732--1745, 2013.
\newblock \doi{10.1093/gji/ggt090}.

\bibitem[Gualtieri et~al.(2014)Gualtieri, Stutzmann, Farra, Capdeville,
  Schimmel, Ardhuin, and Morelli]{Gualtieri2014}
Lucia Gualtieri, El{\'{e}}onore Stutzmann, V.~Farra, Y.~Capdeville,
  M.~Schimmel, Fabrice Ardhuin, and A.~Morelli.
\newblock {Modelling the ocean site effect on seismic noise body waves}.
\newblock \emph{Geophys. J. Int.}, 197\penalty0 (2):\penalty0 1096--1106, 2014.
\newblock \doi{10.1093/gji/ggu042}.

\bibitem[Gutenberg(1947)]{Gutenberg1947}
Beno Gutenberg.
\newblock {Microseisms and weather forecasting}.
\newblock \emph{J. Meteorol.}, 4\penalty0 (1):\penalty0 21--28, 1947.

\bibitem[Hasselmann(1963)]{Hasselmann1963}
K.~Hasselmann.
\newblock {A Statistical Analysis of the Generation of Microseisms}.
\newblock \emph{Rev. Geophys.}, 1\penalty0 (2):\penalty0 177--210, 1963.
\newblock \doi{10.1029/RG001i002p00177}.

\bibitem[Hasselmann et~al.(1976)Hasselmann, Sell, Ross, and
  M{\"{u}}ller]{Hasselmann1976}
K.~Hasselmann, W~Sell, D~B Ross, and P~M{\"{u}}ller.
\newblock {A Parametric Wave Prediction Model}.
\newblock 6\penalty0 (2):\penalty0 200--228, 3 1976.
\newblock \doi{10.1175/1520-0485(1976)006<0200:APWPM>2.0.CO;2}.

\bibitem[Hathi et~al.(2009)Hathi, Ball, Colombatti, Ferri, Leese, Towner,
  Withers, Fulchigioni, and Zarnecki]{Hathi2009}
B.~Hathi, A.~J. Ball, G.~Colombatti, F.~Ferri, M.~R. Leese, M.~C. Towner,
  P.~Withers, M.~Fulchigioni, and J.~C. Zarnecki.
\newblock {Huygens HASI servo accelerometer: A review and lessons learned}.
\newblock \emph{Planet. Space Sci.}, 57\penalty0 (12):\penalty0
  1321--1333, 2009.
\newblock \doi{10.1016/j.pss.2009.06.023}.

\bibitem[Hayes(2016)]{Hayes2016}
Alexander~G. Hayes.
\newblock {The Lakes and Seas of Titan}.
\newblock \emph{Annu. Rev. Earth Planet. Sci.}, 44\penalty0
  (1):\penalty0 57--83, 6 2016.
\newblock \doi{10.1146/annurev-earth-060115-012247}.

\bibitem[Hayes et~al.(2010)Hayes, Wolf, Aharonson, Zebker, Lorenz, Kirk,
  Paillou, Lunine, Wye, Callahan, Wall, and Elachi]{Hayes2010}
Alexander~G. Hayes, A.~S. Wolf, O.~Aharonson, H.~Zebker, Ralph~D. Lorenz, R.~L.
  Kirk, P.~Paillou, J.~Lunine, L.~Wye, P.~Callahan, Stephen~D. Wall, and
  C.~Elachi.
\newblock {Bathymetry and absorptivity of Titan's Ontario Lacus}.
\newblock \emph{J. Geophys. Res. Planets}, 115\penalty0
  (9):\penalty0 1--11, 2010.
\newblock \doi{10.1029/2009JE003557}.

\bibitem[Hayes et~al.(2013)Hayes, Lorenz, Donelan, Manga, Lunine, Schneider,
  Lamb, Mitchell, Fischer, Graves, Tolman, Aharonson, Encrenaz, Ventura,
  Casarano, and Notarnicola]{Hayes2013}
Alexander~G. Hayes, Ralph~D. Lorenz, M.A. Donelan, M.~Manga, J.I. Lunine,
  T.~Schneider, M.P. Lamb, J.M. Mitchell, W.W. Fischer, S.D. Graves, H.L.
  Tolman, O.~Aharonson, P.J. Encrenaz, B.~Ventura, D.~Casarano, and
  C.~Notarnicola.
\newblock {Wind driven capillary-gravity waves on Titan’s lakes: Hard to
  detect or non-existent?}
\newblock \emph{Icarus}, 225\penalty0 (1):\penalty0 403--412, 7 2013.
\newblock \doi{10.1016/j.icarus.2013.04.004}.

\bibitem[Hayes et~al.(2018)Hayes, Lorenz, and Lunine]{Hayes2018}
Alexander~G. Hayes, Ralph~D. Lorenz, and Jonathan~I. Lunine.
\newblock {A post-Cassini view of Titan's methane-based hydrologic cycle}.
\newblock \emph{Nat. Geosci.}, 11\penalty0 (5):\penalty0 306--313, 2018.
\newblock ISSN 17520908.
\newblock \doi{10.1038/s41561-018-0103-y}.

\bibitem[Hofgartner et~al.(2014)Hofgartner, Hayes, Lunine, Zebker, Stiles,
  Sotin, Barnes, Turtle, Baines, Brown, Buratti, Clark, Encrenaz, Kirk,
  Le~Gall, Lopes, Lorenz, Malaska, Mitchell, Nicholson, Paillou, Radebaugh,
  Wall, and Wood]{Hofgartner2014}
J.~D. Hofgartner, Alexander~G. Hayes, J.~I. Lunine, H.~Zebker, B.~W. Stiles,
  Christophe Sotin, J.~W. Barnes, Elizabeth~P. Turtle, K.~H. Baines, R.~H.
  Brown, B.~J. Buratti, R.~N. Clark, P.~Encrenaz, Randolph~L. Kirk, A.~Le~Gall,
  R.~M. Lopes, Ralph~D. Lorenz, M.~J. Malaska, K.~L. Mitchell, P.~D. Nicholson,
  P.~Paillou, J.~Radebaugh, Stephen~D. Wall, and C.~Wood.
\newblock {Transient features in a Titan sea}.
\newblock \emph{Nat. Geosci.}, 7\penalty0 (7):\penalty0 493--496, 6 2014.
\newblock \doi{10.1038/ngeo2190}.

\bibitem[Holthuijsen(2007)]{Holthuijsen2007}
Leo~H. Holthuijsen.
\newblock \emph{{Waves in oceanic and coastal waters}}, volume 9780521860.
\newblock Cambridge University Press, 2007.
\newblock \doi{10.1017/CBO9780511618536}.

\bibitem[Kedar et~al.(2008)Kedar, Longuet-Higgins, Webb, Graham, Clayton, and
  Jones]{Kedar2008a}
Sharon Kedar, M.~S. Longuet-Higgins, Frank~H Webb, N.~Graham, Robert~W.
  Clayton, and C.~Jones.
\newblock {The origin of deep ocean microseisms in the North Atlantic Ocean}.
\newblock \emph{Proc. R. Soc. A Math. Phys. Eng. Sci.}, 464\penalty0 (2091):\penalty0 777--793, 2008.
\newblock \doi{10.1098/rspa.2007.0277}.

\bibitem[Land{\`{e}}s et~al.(2010)Land{\`{e}}s, Hubans, Shapiro, Paul, and
  Campillo]{landes2010}
Matthieu Land{\`{e}}s, Fabien Hubans, Nikolai~M Shapiro, Anne Paul, and Michel
  Campillo.
\newblock {Origin of deep ocean microseisms by using teleseismic body waves}.
\newblock \emph{J. Geophys. Res. Solid Earth}, 115\penalty0
  (B5), 2010.

\bibitem[Lognonn{\'{e}} et~al.(2019)Lognonn{\'{e}}, Banerdt, Giardini, Pike,
  et al.]{Lognonne2019}
Philippe Lognonn{\'{e}}, W.~Bruce Banerdt, D~Giardini, W~T Pike, et al.
\newblock {SEIS: Insight's Seismic Experiment for Internal Structure of Mars}.
\newblock \emph{Space Sci. Rev.}, 215\penalty0 (1):\penalty0 12, 1 2019.
\newblock \doi{10.1007/s11214-018-0574-6}.

\bibitem[Longuet-Higgins(1950)]{Longuet-Higgins1950}
M.~S. Longuet-Higgins.
\newblock {A Theory of the Origin of Microseisms}.
\newblock \emph{Philos. Trans. R. Soc. A Math. Phys. Eng. Sci.}, 243\penalty0
  (857):\penalty0 1--35, sep 1950.
\newblock ISSN 1364-503X.
\newblock \doi{10.1098/rsta.1950.0012}.

\bibitem[Lorenz(2015)]{Lorenz2015}
Ralph~D. Lorenz.
\newblock {Voyage across Ligeia Mare: Mechanics of sailing on the hydrocarbon
  seas of Saturn׳s Moon, Titan}.
\newblock \emph{Ocean Eng.}, 104:\penalty0 119--128, 2015.
\newblock \doi{10.1016/j.oceaneng.2015.04.084}.

\bibitem[Lorenz and Hayes(2012)]{Lorenz2012}
Ralph~D. Lorenz and Alexander~G. Hayes.
\newblock {The growth of wind-waves in Titan's hydrocarbon seas}.
\newblock \emph{Icarus}, 219\penalty0 (1):\penalty0 468--475, 2012.
\newblock \doi{10.1016/j.icarus.2012.03.002}.

\bibitem[Lorenz and Panning(2018)]{Lorenz2017}
Ralph~D. Lorenz and Mark~P. Panning.
\newblock {Empirical recurrence rates for ground motion signals on planetary
  surfaces}.
\newblock \emph{Icarus}, 303:\penalty0 273--279, 3 2018.
\newblock \doi{10.1016/j.icarus.2017.10.008}.

\bibitem[Lorenz et~al.(2010)Lorenz, Newman, and Lunine]{Lorenz2010}
Ralph~D. Lorenz, Claire Newman, and Jonathan~I. Lunine.
\newblock {Threshold of wave generation on Titan's lakes and seas: Effect of
  viscosity and implications for Cassini observations}.
\newblock \emph{Icarus}, 207\penalty0 (2):\penalty0 932--937, 2010.
\newblock \doi{10.1016/j.icarus.2009.12.004}.

\bibitem[Lorenz et~al.(2012)Lorenz, Newman, Tokano, Mitchell, Charnay,
  Lebonnois, and Achterberg]{Lorenz2012c}
Ralph~D. Lorenz, Claire~E. Newman, Tetsuya Tokano, Jonathan~L. Mitchell,
  Benjamin Charnay, Sebastien Lebonnois, and Richard~K. Achterberg.
\newblock {Formulation of a wind specification for Titan late polar summer
  exploration}.
\newblock \emph{Planet. Space Sci.}, 70\penalty0 (1):\penalty0 73--83,
  2012.
\newblock \doi{10.1016/j.pss.2012.05.015}.

\bibitem[Lorenz et~al.(2014)Lorenz, Kirk, Hayes, Anderson, Lunine, Tokano,
  Turtle, Malaska, Soderblom, Lucas, Karatekin, and Wall]{Lorenz2014}
Ralph~D. Lorenz, Randolph~L. Kirk, Alexander~G. Hayes, Yanhua~Z. Anderson,
  Jonathan~I. Lunine, Tetsuya Tokano, Elizabeth~P. Turtle, Michael~J. Malaska,
  Jason~M. Soderblom, Antoine Lucas, {\"{O}}zg{\"{u}}r Karatekin, and Stephen~D. Wall.
\newblock {A radar map of Titan Seas: Tidal dissipation and ocean mixing
  through the throat of Kraken}.
\newblock \emph{Icarus}, 237:\penalty0 9--15, 2014.
\newblock \doi{10.1016/j.icarus.2014.04.005}.

\bibitem[Lorenz et~al.(2018)Lorenz, Turtle, Barnes, Trainer, Adams, Hibbard,
  Sheldon, Zacny, Peplowski, Lawrence, Ravine, Mcgee, Sotzen, Mackenzie,
  Langelaan, Schmitz, Wolfarth, and Bedini]{Dragonfly2018}
Ralph~D. Lorenz, Elizabeth~P. Turtle, Jason~W. Barnes, Melissa~G Trainer,
  Douglas~S Adams, Kenneth~E Hibbard, Colin~Z Sheldon, Kris Zacny, Patrick~N
  Peplowski, David~J Lawrence, Michael~A Ravine, Timothy~G Mcgee, Kristin~S
  Sotzen, Shannon~M Mackenzie, Jack~W Langelaan, Sven Schmitz, Larry~S
  Wolfarth, and Peter~D Bedini.
\newblock {Dragonfly : A Rotorcraft Lander Concept for Scientific Exploration
  at Titan}.
\newblock \emph{Johns Hopkins APL Tech Digest}, 34\penalty0 (4):\penalty0
  374--387, 2018.

\bibitem[Mastrogiuseppe et~al.(2014)Mastrogiuseppe, Poggiali, Hayes, Lorenz,
  Lunine, Picardi, Seu, Flamini, Mitri, Notarnicola, Paillou, and
  Zebker]{Mastrogiuseppe2014}
Marco Mastrogiuseppe, Valerio Poggiali, Alexander~G. Hayes, Ralph~D. Lorenz,
  Jonathan Lunine, Giovanni Picardi, Roberto Seu, Enrico Flamini, Giuseppe
  Mitri, Claudia Notarnicola, Philippe Paillou, and Howard Zebker.
\newblock {The bathymetry of a Titan sea}.
\newblock \emph{Geophys. Res. Lett.}, 41\penalty0 (5):\penalty0
  1432--1437, 3 2014.
\newblock \doi{10.1002/2013GL058618}.

\bibitem[Mitchell et~al.(2015)Mitchell, Barmatz, Jamieson, Lorenz, and
  Lunine]{Mitchell2015}
Karl~L. Mitchell, Martin~B. Barmatz, Corey~S. Jamieson, Ralph~D. Lorenz, and
  Jonathan~I. Lunine.
\newblock {Laboratory measurements of cryogenic liquid alkane microwave
  absorptivity and implications for the composition of Ligeia Mare, Titan}.
\newblock \emph{Geophys. Res. Lett.}, 42\penalty0 (5):\penalty0
  1340--1345, 2015.
\newblock \doi{10.1002/2014GL059475}.

\bibitem[Panning et~al.(2018)Panning, St{\"{a}}hler, Huang, Vance, Kedar, Tsai,
  Pike, and Lorenz]{Panning2018}
Mark~P. Panning, Simon~C. St{\"{a}}hler, Hsin-Hua Huang, Steven~D. Vance,
  Sharon Kedar, Victor~C. Tsai, William~T. Pike, and Ralph~D. Lorenz.
\newblock {Expected Seismicity and the Seismic Noise Environment of Europa}.
\newblock \emph{J. Geophys. Res.: Planets}, 123\penalty0
  (1):\penalty0 163--179, 1 2018.
\newblock \doi{10.1002/2017JE005332}.

\bibitem[Peters et~al.(2012)Peters, Anandakrishnan, Alley, and
  Voigt]{Peters2012}
L.~E. Peters, S.~Anandakrishnan, R.~B. Alley, and D.~E. Voigt.
\newblock {Seismic attenuation in glacial ice: A proxy for englacial
  temperature}.
\newblock \emph{J. Geophys. Res.: Earth Surface}, 117\penalty0
  (2):\penalty0 1--10, 2012.
\newblock \doi{10.1029/2011JF002201}.

\bibitem[Peterson(1993)]{Peterson1993}
Jon Peterson.
\newblock {Observations and Modeling of Seismic Background Noise}.
\newblock Technical report, USGS, Albuquerque, New Mexico, 1993.

\bibitem[Pierson and Moskowitz(1964)]{Pierson1964}
Willard~J. Pierson and Lionel Moskowitz.
\newblock {A proposed spectral form for fully developed wind seas based on the
  similarity theory of S. A. Kitaigorodskii}.
\newblock \emph{J. Geophys. Res.}, 69\penalty0 (24):\penalty0
  5181--5190, 12 1964.
\newblock \doi{10.1029/JZ069i024p05181}.

\bibitem[Pike et~al.(2019)Pike, Lognonn{\'{e}}, Banerdt, Calcutt, Standley,
  Giardini, Charalambous, Stott, McClean, Warren, Zweifel, Mance, Ten~Pierick,
  Mimoun, Murdoch, Hurst, Teanby, Wookey, Myhill, Horleston, Beucler, Clinton,
  Ceylan, van Driel, and Stahler]{Pike2019}
W.~T. Pike, Philippe Lognonn{\'{e}}, W.~B. Banerdt, S.~B. Calcutt, I.~M.
  Standley, D.~Giardini, C.~Charalambous, A.~E. Stott, J.~B. McClean,
  T.~Warren, P.~Zweifel, D.~Mance, J.~Ten~Pierick, D.~Mimoun, N.~Murdoch,
  K.~Hurst, N.~Teanby, J.~Wookey, R.~Myhill, A.~Horleston, E.~Beucler,
  J.~Clinton, S.~Ceylan, M.~van Driel, and S.~C.~St\"{a}hler.
\newblock {Results from the Short-Period (SP) Seismometers on the Mars Insight
  Mission: From Launch to Sol 40}.
\newblock \emph{50th Lunar and Planetary Science Conference}, 50:\penalty0
  2109, 2019.
\newblock URL \url{http://adsabs.harvard.edu/abs/2019LPI....50.2109P}.

\bibitem[Sens-Sch{\"{o}}nfelder and Wegler(2006)]{Sens-Schonfelder2006}
Christoph Sens-Sch{\"{o}}nfelder and Ulrich Wegler.
\newblock {Passive image interferometry and seasonal variations of seismic
  velocities at Merapi Volcano, Indonesia}.
\newblock \emph{Geophysical Research Letters}, 33:\penalty0 1–5, 2006.

\bibitem[Shapiro et~al.(2005)Shapiro, Campillo, Stehly, and
  Ritzwoller]{Shapiro2005}
N.~M. Shapiro, Michel Campillo, Laurent Stehly, and Michael~H. Ritzwoller.
\newblock {High-resolution surface-wave tomography from ambient seismic noise}.
\newblock \emph{Science}, 307\penalty0 (5715):\penalty0 1615, 2005.
\newblock \doi{10.1126/science.1108339}.

\bibitem[Snieder and Safak(2006)]{Snieder2006a}
Roel Snieder and E.~Safak.
\newblock {Extracting the building response using seismic interferometry:
  theory and application to the Millikan Library in Pasadena, California}.
\newblock \emph{Bull. Seismol. Soc. Am.}, 96\penalty0
  (2):\penalty0 586, 2006.
\newblock \doi{10.1785/0120050109}.

\bibitem[Srokosz et~al.(1992)Srokosz, Challenor, and Zarnecki]{Srokosz1992}
M~A Srokosz, P~G Challenor, and J~C Zarnecki.
\newblock \emph{Symposium on Waves on Titan}, 2\penalty0 (2):\penalty0 3--6, 1992.

\bibitem[St{\"{a}}hler et~al.(2018)St{\"{a}}hler, Panning, Vance, Lorenz, van
  Driel, Nissen‐Meyer, and Kedar]{Stahler2018}
Simon~C. St{\"{a}}hler, Mark~P. Panning, Steven~D. Vance, Ralph~D. Lorenz,
  Martin van Driel, Tarje Nissen‐Meyer, and Sharon Kedar.
\newblock {Seismic Wave Propagation in Icy Ocean Worlds}.
\newblock \emph{J. Geophys. Res. Planets}, 123\penalty0
  (1):\penalty0 206--232, 1 2018.
\newblock \doi{10.1002/2017JE005338}.

\bibitem[Stutzmann et~al.(2009)Stutzmann, Schimmel, Patau, and
  Maggi]{Stutzmann2009}
Eléonore Stutzmann, Martin Schimmel, Geneviève Patau, and Alessia Maggi.
\newblock {Global climate imprint on seismic noise}.
\newblock \emph{Geochemistry, Geophysics, Geosystems}, 10\penalty0
  (11):\penalty0 n/a--n/a, 11 2009.
\newblock \doi{10.1029/2009GC002619}.

\bibitem[Tokano(2013)]{Tokano2013}
Tetsuya Tokano.
\newblock {Are tropical cyclones possible over Titan’s polar seas?}
\newblock \emph{Icarus}, 223\penalty0 (2):\penalty0 766--774, 4 2013.
\newblock \doi{10.1016/j.icarus.2013.01.023}.

\bibitem[Turtle et~al.(2011)Turtle, Perry, Hayes, Lorenz, Barnes, McEwen, West,
  Del~Genio, Barbara, Lunine, Schaller, Ray, Lopes, and Stofan]{Turtle2011}
Elizabeth~P. Turtle, J~E Perry, Alexander~G. Hayes, Ralph~D. Lorenz, J~W
  Barnes, A.~S. McEwen, R~A West, A.~D. Del~Genio, J~M Barbara, J~I Lunine, E~L
  Schaller, T~L Ray, R~M~C Lopes, and E.~R. Stofan.
\newblock {Rapid and Extensive Surface Changes Near Titan's Equator: Evidence
  of April Showers}.
\newblock \emph{Science}, 331\penalty0 (6023):\penalty0 1414--1417, 3 2011.
\newblock \doi{10.1126/science.1201063}.

\bibitem[Vance et~al.(2018{\natexlab{a}})Vance, Kedar, Panning, St{\"{a}}hler,
  Bills, Lorenz, Huang, Pike, Castillo, Lognonn{\'{e}}, Tsai, and
  Rhoden]{Vance2018}
Steven~D. Vance, Sharon Kedar, Mark~P. Panning, Simon~C. St{\"{a}}hler, Bruce~G
  Bills, Ralph~D. Lorenz, Hsin-hua Huang, William~T. Pike, Julie~C Castillo,
  Philippe Lognonn{\'{e}}, Victor~C. Tsai, and Alyssa~R. Rhoden.
\newblock {Vital Signs: Seismology of Icy Ocean Worlds}.
\newblock \emph{Astrobiology}, 18\penalty0 (1):\penalty0 37--53, 1
  2018{\natexlab{a}}.
\newblock \doi{10.1089/ast.2016.1612}.

\bibitem[Vance et~al.(2018{\natexlab{b}})Vance, Panning, St{\"{a}}hler,
  Cammarano, Bills, Tobie, Kamata, Kedar, Sotin, Pike, Lorenz, Huang, Jackson,
  and Banerdt]{Vance2018a}
Steven~D. Vance, Mark~P. Panning, Simon~C. St{\"{a}}hler, Fabio Cammarano,
  Bruce~G. Bills, Gabriel Tobie, Shunichi Kamata, Sharon Kedar, Christophe
  Sotin, William~T. Pike, Ralph~D. Lorenz, Hsin-Hua Huang, Jennifer~M. Jackson,
  and W.~Bruce Banerdt.
\newblock {Geophysical Investigations of Habitability in Ice-Covered Ocean
  Worlds}.
\newblock \emph{J. Geophys. Res. Planets}, 123\penalty0
  (1):\penalty0 180--205, 1 2018{\natexlab{b}}.
\newblock \doi{10.1002/2017JE005341}.

\bibitem[Wall et~al.(2010)Wall, Hayes, Bristow, Lorenz, Stofan, Lunine,
  Le~Gall, Janssen, Lopes, Wye, Soderblom, Paillou, Aharonson, Zebker, Farr,
  Mitri, Kirk, Mitchell, Notarnicola, Casarano, and Ventura]{Wall2010}
Stephen~D. Wall, Alexander~G. Hayes, C.~Bristow, Ralph~D. Lorenz, E.~R. Stofan,
  J.~Lunine, A.~Le~Gall, Mike Janssen, R.~Lopes, L.~Wye, L.~Soderblom,
  P.~Paillou, O.~Aharonson, H.~Zebker, T.~Farr, G.~Mitri, R.~Kirk, K.~Mitchell,
  C.~Notarnicola, D.~Casarano, and B.~Ventura.
\newblock {Active shoreline of Ontario Lacus, Titan: A morphological study of
  the lake and its surroundings}.
\newblock \emph{Geophys. Res. Lett.}, 37\penalty0 (5):\penalty0
  n/a--n/a, 3 2010.
\newblock \doi{10.1029/2009GL041821}.

\bibitem[Younglove and Ely(1987)]{Younglove1987}
B.~A. Younglove and J.~F. Ely.
\newblock {Thermophysical Properties of Fluids. II. Methane, Ethane, Propane,
  Isobutane, and Normal Butane}.
\newblock \emph{J. Phys. Chem. Ref. Data}, 16\penalty0
  (4):\penalty0 577--798, 10 1987.
\newblock \doi{10.1063/1.555785}.

\end{thebibliography}
\end{document}